\documentclass{elsart}
\usepackage{natbib}
\usepackage{amsmath}
\usepackage{amssymb}
\usepackage{amsfonts}
\usepackage{graphicx}
\usepackage{mathrsfs}
\journal{Stud. Hist. Phil. M. P.}

\begin{document}
\newtheorem{theo}{Theorem}
\newtheorem{definition}{Definition}
\begin{frontmatter}

\title{Mechanical Proof of the Second Law of Thermodynamics Based on Volume Entropy}
\author{Michele Campisi}
%\thanks[aut]{}
\address{Department of Physics,University of North Texas, P.O.
Box 311427, Denton, TX 76203-1427,USA} \ead{campisi@unt.edu}

\begin{abstract}
In a previous work (M. Campisi. Stud. Hist. Phil. M. P. 36 (2005)
275-290) we have addressed the mechanical foundations of
equilibrium thermodynamics on the basis of the Generalized
Helmholtz Theorem. It was found  that the volume entropy provides
a good mechanical analogue of thermodynamic entropy because it
satisfies the heat theorem and it is an adiabatic invariant. This
property explains the ``equal'' sign in Clausius principle ($S_f
\geq S_i$) in a purely mechanical way and suggests that the volume
entropy might explain the ``larger than'' sign (i.e. the Law of
Entropy Increase) if non adiabatic transformations were
considered. Based on the principles of microscopic (quantum or
classical) mechanics here we prove that, provided the initial
equilibrium satisfy the natural condition of decreasing ordering
of probabilities, the expectation value of the volume entropy
cannot decrease for arbitrary transformations performed by some
external sources of work on a insulated system. This can be
regarded as a rigorous quantum mechanical proof of the Second Law.
We discuss how this result relates to the Minimal Work Principle
and improves over previous attempts. The natural evolution of
entropy is towards larger values because the natural state of
matter is at positive temperature. Actually the Law of Entropy
Decrease holds in artificially prepared negative temperature
systems.
\end{abstract}

\end{frontmatter}
\section{\label{sec:introduction}Introduction}
This work addresses the problem of explaining the Second Law of
Thermodynamics on the basis of the microscopic laws of mechanics.
As discussed earlier \citep{Campisi05} the Second Law of
thermodynamics is commonly understood as composed of two parts
which we shall conventionally label as ``part A'' and ``part B''.
``Part A'' is essentially a statement about the \emph{Existence of
Entropy}. It says that there exists an integrating factor
$\frac{1}{T}$, interpreted as inverse temperature, such that
$\frac{\delta Q}{T}$ is an exact differential, where $\delta
Q=dE+PdV$ is the heat exchanged during a very small variation of
the thermodynamic state $(E,V)$. This implies that there exists a
function of the thermodynamic state, the entropy $S$, that
generates the exact differential
\begin{equation}\label{}
    dS = \frac{\delta Q}{T}.
\end{equation}
``Part A'' of the second Law evidently pertains to
\emph{equilibrium thermodynamics}. Its mechanical foundations have
been studied in a previous work \citep{Campisi05}. The main
conclusion drawn in that work was that the laws of ergodic
Hamiltonian mechanics alone are sufficient for providing a rather
satisfactory explanation of this part of the Second Law. The
approach adopted was that of establishing a correspondence between
the \emph{thermodynamic} quantities $E,P,T,V$ and certain suitably
chosen time-averaged \emph{mechanical} quantities. Thanks to a
generalization of Helmholtz Theorem, then we were able to find the
proper \emph{mechanical analogue} of thermodynamic entropy. This
is the so called \emph{Volume Entropy}:
\begin{equation}\label{eq:volumeEntropy}
    S_\Phi(E,V) = \ln \int_{H(\mathbf{q},\mathbf{p};V)\leq E} \frac{d^{3N}\mathbf{q}
    d^{3N}\mathbf{p}}{h^{3N}}
\end{equation}
where $H(\mathbf{q},\mathbf{p};V)$ is the system's Hamilton
function. This is a function  of the $6N$ dimensional phase-space
vector $\mathbf{q},\mathbf{p}$ and the external parameter $V$. $h$
is an arbitrary constant with the dimensions of action.

The present contribution completes the previous one by addressing
``Part B'' of the Second Law. ``Part B'' is essentially the Law of
Entropy increase, and as such is a statement that pertains to
\emph{non-equilibrium thermodynamics}. To avoid any possible
confusion, here by ``Part B'' of the Second Law and ``Law of
Entropy increase'', we mean Clausius formulation of the Entropy
Principle \citep{Uffink01}:
\begin{quote}
THE ENTROPY PRINCIPLE: For every \emph{nicht umkehrbar} process in
a thermally isolated system which begins and ends in an
equilibrium state, the entropy of the final state is greater than
or equal to that of the initial state. For every \emph{umkehrbar}
process in a thermally isolated system the entropy of the final
state is equal to that of the initial state
\end{quote}
The expressions \emph{nicht umkehrbar} and \emph{umkehrbar} could
be translated into the current scientific English as \emph{non
quasi static} and \emph{quasi static} respectively. It must be
stressed that the Entropy Principle refers to transformations
caused by the variation of some external field, and is not at all
a statement about the \emph{spontaneous tendencies} of physical
systems. If the variation of the external field is acted in such a
way as to drive the system out of equilibrium (non quasi static
process) the entropy will increase. If it is acted in such a way
that the system will remain arbitrary close to equilibrium (quasi
static process) then the entropy will not change.

Indeed the volume entropy already well accounts for the ``quasi
static'' part of the Entropy Principle. In facts, as it is known
since the work of \cite{Hertz}, the volume entropy is an
\emph{ergodic adiabatic invariant}.  Namely a quantity that does
not change during a quasi-static transformation of the external
field. As pointed out in \citep{Campisi05}, the laws of ergodic
Hamiltonian mechanics alone provide a quite satisfactory
explanation of the ``quasi static'' part of the Entropy Principle:
no statistics is needed to explain the equal sign in the Entropy
Principle. It must be emphasized that here we are establishing a
correspondence between the thermodynamic concept of quasi static
process and the classical mechanical concept of adiabatic
transformation. In particular the expression ``adiabatic'' will
not be used as synonym of ``thermally insulated'' as customarily
happens in Thermodynamics text-books.

Obviously the fact that the Volume Entropy well accounts for
``part A'' and the quasi static part of ``part B'' of the Second
Law, suggests that it might turn out to be very useful in
addressing the non quasi static part of ``part B''. Simple
considerations suggest that the latter could be proved only in
some statistical or averaged sense, though. Consider for example a
1D particle of mass $m$ in a 1D box of length $L$. The particle
bounces forth and back inside the box. Let $E$ be the energy of
the particle. Imagine that we can change to length of the box by
moving the right wall of the box. The volume entropy of this
elementary ergodic system is simply $S_\Phi(E,L) =
\ln(2L\sqrt{2mE})$. Now imagine that we perform a very fast
compression of the box, much faster than the particle period of
motion $T = \sqrt{\frac{m}{2E}}L$. Let $L-\Delta L$ be the final
length of the box. Imagine that during this transformation the
particle is far from the moving wall and does not bounce against
it. Its energy would not change but the change of its volume
entropy would be $\Delta S_\Phi = \ln (1-\Delta L/L)$. Namely it
would be negative. This simple argument should convince that any
purely mechanical attempt to prove the Entropy Principle on the
basis of Volume Entropy would be vain. We certainly need to add
some statistical ingredient if we want to prove it.

Thus we are going to assume that the initial energy of our
insulated system is not known. All we know is that it is within
some range $E, E+dE$ with some probability $p(E)\Omega(E)dE$. The
symbol $\Omega(E)$ denotes the ``density of states'' at energy
$E$. $\Omega(E)$ is also named surface integral \citep{Campisi05}
or structure function \citep{Khinchin}. For example, if we first
place the system in thermal contact with a heat bath at
temperature $T$, and then we remove the contact, we will not know
for certain what the energy of the system will be, but we will
know that $p(E) = \frac{e^{-E/T}}{Z}$.

In this work we will prove that, provided that $p(E)$ is a
decreasing function of $E$, the \emph{expectation value} of the
Volume Entropy will be larger than its initial one. It turns out
that such proof is much easier in quantum mechanics rather than
classical mechanics. Therefore we shall first quantize the Volume
Entropy and then study its behavior under the action of a varying
field, that is a time-dependent perturbation.

The paper is organized as follows. In Sec. \ref{sec:QVE} we
introduce the quantum counterpart of the classical volume entropy.
In Sec. \ref{sec:perturbation} we prove that the expectation value
of such quantum operator can only increase, if the initial
conditions is represented by a decreasing ordering of
probabilities. In Sec. \ref{sec:kelvin} we discuss how this result
relates to Thomson's formulation of the second Law, whereas in
Sec. \ref{sec:comparison} we compare the quantum volume entropy
with other quantum entropies present in the literature. In Sec.
\ref{sec:Classical Case} we show how to adapt the quantal proof of
Sec. \ref{sec:perturbation} to the classical case. The role of the
initial equilibrium is discussed in Sec.
\ref{sec:initial-condition}. We will discuss the fact that the
results proven in the paper are direct consequences of the
time-reversal symmetric microscopic laws, and that, besides the
law of entropy increase, there exists a law of entropy decrease as
well. The conclusions are drawn in Sec. \ref{sec:conclusion}.

\section{\label{sec:QVE}Quantum Volume Entropy}
Let us first consider the 1D case. In 1D the volume entropy reads:
\begin{equation}
S_\Phi =\ln \int_{H \leq E} \frac{dxdp}{h}. \label{}
\end{equation}
This can be conveniently reexpressed as the logarithm of the
\emph{reduced action} \citep{Landau1}
\begin{equation}
S_\Phi =\ln \oint \frac{pdx}{h}. \label{eq:helmEnt}
\end{equation}
This is also known as the Helmholtz Entropy \citep{Campisi05}.
Quantization of the Helmholtz Entropy is almost immediate. Indeed,
using a colorful expression, I would say that Eq.
(\ref{eq:helmEnt}) invites the reader to quantize. Using the
semiclassical approximation of Bohr-Sommerfeld \citep{Landau3} and
setting $h$ equal to Plank's constant allows to see that $S_\Phi$
is a quantized quantity whose possible values are (within the
range of validity of the approximation):
\begin{equation}\label{eq:S=logn}
    S_\Phi = \ln \left(n+\frac{1}{2}\right)
\end{equation}
We can extend this reasoning to multidimensional systems whose
dynamics is ergodic. In this general case the Volume Entropy is
given by Eq. (\ref{eq:volumeEntropy}). Again using the
quasi-classical viewpoint \citep{Landau5} the integral in Eq.
(\ref{eq:volumeEntropy}) approximately counts the number of
quantum states not above a certain energy $\varepsilon_n = E$.
Since the levels are non degenerate this number is
$n+\frac{1}{2}$, where one considers that the vacuum state counts
as a half state. The levels are non-degenerate because the
corresponding classical dynamics is ergodic. This can be
understood by noticing that ergodicity implies that the
Hamiltonian is the only integral of motion. This, translated into
the language of quantum mechanics, says that the Hamiltonian alone
constitute a complete set of commuting observables, so that the
only quantum number is $n$.

At this point, it is quite easy to construct the quantum version of
Volume Entropy. Consider a finite (i.e., not necessarily
macroscopic) \emph{non-degenerate} quantum systems. Let
$\mathcal{N}$ be the \emph{quantum number operator}, i.e.:
\begin{equation}\label{}
    \mathcal{N} \doteq \sum_{k=0}^K k |k\rangle\langle k|
\end{equation}
where $\{|k\rangle\}$ is the complete orthonormal set of
Hamiltonian's eigenstates. $K$, the total number of energy levels,
can be infinite. The eigenvectors of $\mathcal{N}$ are the energy
eingenvectors, and the eigenvalues are the corresponding quantum
numbers. Then the \emph{Quantum Volume Entropy Operator} can be
defined as:
\begin{equation}\label{eq:S=logN}
    \mathcal{S} \doteq \ln \left(\mathcal{N}+\frac{1}{2}\right) %\left(\mathcal{N} + \frac{1}{2} \right)
\end{equation}
We adopt a system of units where $k_B$, Boltzmann constant, is equal
to $1$.

\section{\label{sec:perturbation}Proof of the Entropy Principle}
Armed with a quanto-mechanical analogue of thermodynamic entropy
(\ref{eq:S=logN}), we can now study its evolution under a
time-dependent perturbation. As prescribed by the Entropy
Principle we shall assume that the system is thermally isolated
from the environment. As discussed previously, the system energy
is not known. This means that the system is assumed to be in a
statistical mixture of states, described by a density matrix
$\rho_i$, rather than a pure state $|k\rangle$. As prescribed by
the Entropy Principle we shall also assume that the system be
initially at equilibrium. We will translate this thermodynamic
notion into the quantal requirement that $ \frac{\partial
\rho_i}{\partial t} = 0 $. So the system is at
 equilibrium whenever $\frac{\partial \rho}{\partial t} = 0 $ and it is
out of equilibrium whenever $\frac{\partial \rho}{\partial t} \neq
0 $. At $t=t_i$, we switch on a perturbation. This is implemented
by changing the value of some external parameter $\lambda$ during
the course of time: $\lambda = \lambda(t)$. $\lambda$ can be for
example the volume $V$ of a vessel containing the system, or the
value of some external field like an electric or a magnetic field.
At time $t=t_{off}$, the perturbation is switched off. We assume
that at some time $t_f \geq t_{off}$ any transient effect will be
vanished and the system attains a new equilibrium state described
by some $\rho_f$, such that $\frac{\partial \rho_f}{\partial t} =
0 $. Thus, before time $t_i$ and after $t_{f}$, the system is at
equilibrium, and for $t_i<t<t_f$ it is out of equilibrium. Due to
the perturbation the Hamiltonian changes from the initial value
$H_i$ to the final value $H_f$, and accordingly the quantum
entropy operator will change in time and move from $\mathcal{S}_i$
to $\mathcal{S}_f$. We introduce the following time-dependent
orthonormal basis set $\{|k,t\rangle\}$. The vectors $|k,t\rangle$
are defined as the eigenvectors of the ``frozen'' Hamiltonian
$\overline{H}_t \doteq H(t)$. That is:
\begin{equation}\label{eq:spectum-H}
    H(t) = \sum_{k=0}^K \varepsilon_k(t)|k,t\rangle \langle k,t|
\end{equation}
Since at time $t_i$ $\frac{\partial \rho_i}{\partial t} = 0$, then
$[\rho_i,H]=0$. This means that $\rho_i$ is diagonal over the
initial basis $\{|k,t_i\rangle\}$:
\begin{equation}\label{}
    \rho(t_i) = \sum_{k=0}^K p_k |k,t_i\rangle \langle k,t_i|
\end{equation}
As anticipated in the introduction we shall assume that $p_i$ is
decreasing:
\begin{equation}\label{eq:orderingProbabilities}
    p_0 \geq p_1 \geq ... \geq p_i \geq...
\end{equation}
Our definition of quantum entropy (\ref{eq:S=logN}), is essentially
an \emph{equilibrium} definition. Using the bases $\{|k,t\rangle\}$,
we can extend the definition to the \emph{out of equilibrium} case,
as following:
\begin{equation}\label{eq:spectrum-S(t)}
    \mathcal{S}(t) \doteq  \sum_{k=0}^K \ln \left(k+\frac{1}{2}\right) |k,t\rangle \langle k,t|
\end{equation}
We shall assume that non-degeneracy is kept at all times. This
implies that there is \emph{no level crossing}, and ensures that
the quantum number operator gives the correct eigenvalues at all
times. The same assumption is used by \citet{Allahverdyan05} to
ensure the proper ordering of energy eigenvalues. Note that,
unlike the Hamiltonian's spectrum (\ref{eq:spectum-H}), the
spectrum of the quantum entropy (\ref{eq:spectrum-S(t)}) is
time-independent. We define the transition probabilities:
\begin{equation}\label{}
    |a_{kn}(t_f)|^2 = |\langle n,t_f|U(t_i,t_f)|k,t_i\rangle  |^2
\end{equation}
Where
\begin{equation}\label{}
    U(t_i,t) = \mathscr{T}\exp \left( -\frac{i}{\hslash}\int_{t_i}^{t} H(s)ds
\right)
\end{equation}
is the time evolution operator expressed in terms of the
time-ordered exponential $\mathscr{T}\exp $. The
$|a_{kn}(t_f)|^2$'s represent the probabilities that the system
will be found in the state $|n,t_f\rangle$ at time $t_f$ provided
that it was in the state $|k,t_i\rangle$ at time $t_i$. They
satisfy the relations \citep{Allahverdyan05}:
\begin{equation}\label{eq:sum-a-nk}
    \sum_{k=0}^K |a_{kn}(t_f)|^2 = \sum_{n=0}^K |a_{kn}(t_f)|^2 =
    1
\end{equation}
and
\begin{equation}\label{eq:a-nk>0}
    |a_{kn}(t_f)|^2 \geq 0
\end{equation}
For the change in the expectation value of the quantum entropy
$\mathcal{S}$ of Eq. (\ref{eq:S=logN}) we have:
\begin{equation}\label{eq:Sf-si}
    S_f -S_i = Tr \left[\rho_f \mathcal{S}_f \right]- Tr \left[\rho_i
    \mathcal{S}_i\right] =  \sum_{n=0}^K (p'_n-p_n) \ln \left(n+\frac{1}{2}\right)
\end{equation}
where
\begin{equation}\label{eq:p'n}
p'_n= \sum_{k=0}^K p_k|a_{kn}(t_f)|^2
\end{equation}
is the probability that the system is in state $|n,t_f\rangle$
provided that the initial probabilities were $p_n$.

Using the ``summation by parts'' rule \citep{Allahverdyan05}:
\begin{equation}\label{}
    \sum_{n=0}^K a_n b_n  = a_K \sum_{n=0}^K b_n - \sum_{m=0}^{K-1} (a_{m+1}- a_m)\sum_{n=0}^m b_n
\end{equation}
Eq. (\ref{eq:Sf-si}) becomes
\begin{equation}\label{eq:DeltaS}
    S_f -S_i = \sum_{m=0}^K \ln \frac{m+\frac{3}{2}}{m+\frac{1}{2}} \sum_{n=0}^m (p_n-p'_n)
\end{equation}
We have:
\begin{eqnarray}\label{}
    \sum_{n=0}^m (p_n-p'_n)&=& \sum_{n=0}^m
p_n- \sum_{n=0}^m \sum_{i=0}^{K} p_i
|a_{in}(t_f)|^2 \nonumber\\
&=& \sum_{n=0}^m p_n \left(1-\sum_{i=0}^{m} |a_{in}(t_f)|^2
\right) - \sum_{n=0}^m \sum_{i=m+1}^{K} p_i |a_{in}(t_f)|^2
\end{eqnarray}
From Eq.s (\ref{eq:sum-a-nk}) and (\ref{eq:a-nk>0}) we have
$\left(1-\sum_{i=0}^{m} |a_{in}(t_f)|^2 \right) \geq 0$ and
$|a_{in}(t_f)|^2\geq 0$, therefore using the ordering of
probabilities (\ref{eq:orderingProbabilities}) we get (see also
\citep{Allahverdyan05}):
\begin{eqnarray}\label{}
    \sum_{n=0}^m (p_n-p'_n) &\geq& p_m \sum_{n=0}^m \left(1-\sum_{i=0}^{m}
|a_{in}(t_f)|^2 \right) - p_m \sum_{n=0}^m
\sum_{i=m+1}^{K}  |a_{in}(t_f)|^2\\
&=& m p_m - p_m \sum_{n=0}^m \sum_{i=0}^{K} |a_{in}(t_f)|^2 = 0
\end{eqnarray}
where we used Eq. (\ref{eq:sum-a-nk}) in the last line. Noting
that $\ln \frac{m+\frac{3}{2}}{m+\frac{1}{2}} > 0$ in Eq.
(\ref{eq:DeltaS}) , we finally reach the conclusion that:
\begin{equation}\label{eq:Entropy-Increase}
    S_f \geq S_i
\end{equation}
This inequality holds for any transformation acted on a thermally
insulated, non degenerate quantum system which is initially at
equilibrium with a decreasing ordering of probabilities. To
complete the proof of the Entropy Principle we have to prove that
the equal sign holds for adiabatic transformation. Note that the
non-degeneracy assumption ensures that the \emph{quantum adiabatic
theorem} holds \citep{Messiah62}. This ensures that the transition
probability between states with different quantum number will be
null during an adiabatic transformation:
\begin{equation}\label{eq:a-kn=delta-kn}
    |a_{in}(t_f)|^2= \delta_{in}
\end{equation}
Therefore, for an adiabatic transformation we get $p'_i=p_i$ (see
Eq. \ref{eq:p'n}), which brings to
\begin{equation}\label{eq:s-f=s-i}
    S_f = S_i
\end{equation}
This concludes our quanto-mechanical proof of the Entropy
Principle. Note that the result in Eq. (\ref{eq:s-f=s-i}) is not
surprising because the quantum entropy operator has been defined
as the quantum counterpart of a classical adiabatic invariant.

Also note that we have established and used the following
correspondences between thermodynamics and quantum mechanics:
\begin{itemize}
    \item entropy  $\rightleftharpoons \ln\left(\mathcal{N}+\frac{1}{2}\right)$
    \item equilibrium $\rightleftharpoons\frac{\partial \rho}{\partial t} = 0 $
    \item (non)quasi-static process $\rightleftharpoons$(non)adiabatic perturbation
\end{itemize}

Since the equilibrium condition for the final state has never been
used in the proof, inequality (\ref{eq:Entropy-Increase}) holds
for any $t \geq t_i$. Note that this by no means implies that
\begin{equation}\label{eq:non-eq-entropy}
    S(t) \doteq Tr [\mathcal{S}(t)\rho(t)]
\end{equation}
is a monotonic increasing function of time. All we can say is that
if at times $t_1 < t_2 <...<t_n<...$ the density matrix is
diagonal and its spectrum is monotonic decreasing, then:
\begin{equation}\label{}
    S(t_1) \leq S(t_2) \leq ... \leq S(t_n) \leq...
\end{equation}
In general there can well be two times $t_A<t_B$, where for
example the system is out of equilibrium, such that
$S(t_A)>S(t_B)$. It is important to stress that, when the system
is out of equilibrium the quantity $S(t)$ shouldn't be regarded as
the system's thermodynamic entropy, which is essentially an
equilibrium property. Thus $S(t)$ is only one of the many possible
out of equilibrium generalizations of entropy. What makes it
special is that it proves effective in addressing the Entropy
Principle.
\section{\label{sec:kelvin}Thomson's formulation and the Minimal Work Principle}
The present proof of The Entropy Principle is very close, in the
approach and methods, to a result discussed recently by
\citet{Allahverdyan02}. They considered the following alternative
formulation of the Second Law, which they attribute to Kelvin (W.
Thomson):
\begin{quote}
THOMSON'S FORMULATION: No work can be extracted from a closed
equilibrium system during a cyclic variation of a parameter by an
external source.
\end{quote}
If we denote the work done by the external source as $W$, the
principle can be expressed simply as:
\begin{equation}\label{eq:W=0}
    W \geq 0
\end{equation}
The proof of \cite{Allahverdyan02} goes like the one we have
proposed above for the Entropy Principle. Indeed that work has
been a major source of inspiration for the present one. In this
case one wants to study the following quantity:
\begin{equation}\label{}
    W \doteq Tr [H_f\rho_f] - Tr [H_i\rho_i]
\end{equation}
for a cyclic process. This means that the final Hamiltonian is
assumed to be equal to the initial one $H_f=H_i \doteq H_0$. Thus:
\begin{equation}\label{}
    W \doteq \sum_{n=0}^K \varepsilon_n(p'_n-p_n)
\end{equation}
where $\varepsilon_n$ are the eigenvalues of $H_0$. These are
ordered according to $\varepsilon_1 > \varepsilon_2
>...>\varepsilon_i>... $ . The eigenvalues $\varepsilon_n$, play here the same role as the
entropy eigenvalues $\ln(n+1/2)$, in Eq. (\ref{eq:Sf-si}). Thus it
is immediate to see that, under the assumption of decreasing
probabilities (\ref{eq:orderingProbabilities}), Eq. (\ref{eq:W=0})
holds quanto-mechanically.

In a subsequent work \cite{Allahverdyan05} have extended this
result to the case of possibly non cyclic transformation. They
have found that
\begin{equation}\label{eq:W>TildaW}
    W - \widetilde{W} = \sum_{n=0}^K \varepsilon'_n(p'_n-p_n) \geq
    0
\end{equation}
Where $\varepsilon'_n$ are the eigenvalues of the final
Hamiltonian, $W$ is the work actually performed on the system and
$\widetilde{W}$ is the work that would have been performed if the
same transformation would have been carried adiabatically. The
proof is formally equivalent to the one discussed here. Eq.
(\ref{eq:W>TildaW}) expresses the Minimal Work Principle according
to which, whenever we perform a non-adiabatic transformation, we
spend more work than we would have if performing an adiabatic one.

The formal similarity of Eq. (\ref{eq:W>TildaW}) and Eq.
(\ref{eq:Sf-si}) proves that the formulation of the Second Law as
a Principle of  Minimal Work or as an Entropy Principle are
\emph{equivalent}. In particular it is easily seen that:
\begin{equation}\label{}
    \texttt{sign}(W-\widetilde{W}) = \texttt{sign}(S_f-S_i)
\end{equation}
Thus the two principles are equivalent. Further, whenever one is
violated, the other will be too. Cases where the Minimal Work
Principle is violated because of level crossing are discussed by
\cite{Allahverdyan05}. In those case The Entropy Principle would
be violated too.

The history behind this kind of quanto-mechanical proofs of the
Second Law is relatively recent, and can traced back at lest to
the works of \cite{Lenard78} and \cite{Basset78}. Due to the lack
of a suitable quanto-mechanical analogue of entropy, though, the
application of such quantal approaches has remained restricted to
the analysis of statements that concern work, whose mechanical
definition is quite straightforward. To the best of the author's
knowledge, similar arguments and approaches have been previously
proposed for addressing the Entropy Principle only in the
relatively un-known work of \citet{Tasaki00}.
\citeauthor{Tasaki00} already proposed the quantum entropy
operator in the form $\mathcal{S} = \ln \mathcal{N}$, but the
connection with the Generalized Helmholtz Theorem (which has been
introduced later \citep{Campisi05}) was not made, neither the
importance of the volume entropy as a good mechanical analogue of
thermodynamic entropy for possibly low dimensional systems was
recognized. Unlike the present work, in fact, the work of
\citet{Tasaki00} is concerned only with the macroscopic
case.\footnote{The work of \citet{Tasaki00} contained a
simultaneous proof of both Eq. (\ref{eq:Entropy-Increase}) and Eq.
(\ref{eq:W>TildaW}). It is interesting to notice that
\citeauthor{Tasaki00} did not published his result because Eq.
(\ref{eq:W>TildaW}) was proven previously by \citet{Lenard78}. To
the best of my knowledge, the proof of Eq.
(\ref{eq:Entropy-Increase}) based on the logarithm of the
principal quantum number was never given before though, thus it
remained unpublished.}

\section{\label{sec:comparison}Comparison with other quantum entropies}
The employment of the Quantum Volume Entropy improves quite a lot
over previous attempts at explaining the Entropy Principle based
on quantum entropies. In fact, the employment of the entropy in
Eq. (\ref{eq:non-eq-entropy}) has many advantages over other
quantum mechanical entropies present in literature. In contrast
with von Neumann entropy:
\begin{equation}\label{eq:SvN}
    S_{vN} = - Tr [\rho(t) \ln \rho(t)].
\end{equation}
the expectation value of the quantum operator $\mathcal{S}$ does
change in time, and it has been proved to increase under the
assumption discussed. Tolman's coarse-grained entropy
\citep{Tolman38}:
\begin{equation}\label{eq:cgEnt}
    S_{cg}(t) = - \sum_{\nu} P_{\nu}(t) \ln P_{\nu}(t)
\end{equation}
does change in time and it is an adiabatic invariant
\citep{Tolman38}. Nonetheless it fails in accounting for the
inequality sign in the case of non-adiabatic perturbations. All we
known is that for an infinitesimal \emph{abrupt} transformation
that begins and ends in a \emph{canonical equilibrium} the
corresponding change in $S_{cg}$ is non-negative \citep{Tolman38}.
But this does not ensure that for any \emph{finite} non-adiabatic
transformation the change would be non-negative as required by
Clausius formulation. Tolman's argument according to which any
finite transformation could be reproduced by a sequence of many
infinitesimal abrupt transformations each followed by the reaching
of a canonical equilibrium does not seem to be tenable. In fact,
as a result of a finite non-adiabatic transformation, the system
could well end up in a non-canonical distribution
\citep{Allahverdyan05}. Further, Tolman's definition of entropy of
Eq. (\ref{eq:cgEnt}) applies only to macroscopic equilibrium
systems. On the contrary the result proved in Sec.
\ref{sec:perturbation} holds \emph{no matter the number of degrees
of freedom} of the system. Thus the Quantum Volume Entropy might
turn out to be very useful in the novel and fast growing field of
\emph{Quantum Thermodynamics} of nanoscale systems. See for
example \citep{Frontiers05}, see also \citep{Kieu04}.

\section{\label{sec:Classical Case}Classical case}
The result of Eq. (\ref{eq:Entropy-Increase}) can be proved also
classically. Let the system be initially distributed according to
some probability distribution function $p_0(E)$. Let $p_1(E)$ be
the final distribution. Let $\Phi_0(E)$ and $\Phi_1(E)$ denote the
volumes enclosed by the hyper-surfaces $H_0
(\mathbf{q},\mathbf{p}) = E$ and $H_1(\mathbf{q},\mathbf{p}) = E$
respectively. Where $H_0$ and $H_1$ are the initial and final
Hamiltonians. The volume entropy of a representative point that at
time $t_i$ lyes on the hyper-surfaces $H_0 (\mathbf{q},\mathbf{p})
= E$ is $\ln \Phi_0(E)$. We have a similar expression for time
$t_f$. Then:
\begin{equation}\label{}
    S_f-S_i = \int_0^\infty dE \Omega_1(E)p_1(E)\ln \Phi_1(E) - \int_0^\infty dE \Omega_0(E)p_0(E)\ln \Phi_0(E)
\end{equation}
where $\Omega_r$ denotes the initial ($r=0$) or final ($r=1$),
structure function. Note that in general
\begin{equation}\label{}
    dE \Omega_r(E) = d\Phi_r(E)
\end{equation}
Thus we can make the  change of variable $E \leftrightarrow
\Phi_r$ in the integrals. Let $P_r(\Phi_r)\doteq p_r(E(\Phi_r))$,
then we have (after dropping the subscript in $\Phi_r$):
\begin{equation}\label{}
    S_f-S_i = \int_0^\infty d\Phi ( P_1(\Phi) - P_0(\Phi) ) \ln \Phi
\end{equation}
This is the classical analogue of Eq. (\ref{eq:Sf-si}). The role
of $n+\frac{1}{2}$ is played by the ``enclosed volume'' $\Phi$,
and the discrete probabilities $p_n, p'_n$ are now probability
density functions $P_r(\Phi)$. Since the evolution is
deterministic, it is possible to express the final probability in
terms of the initial one as
\begin{equation}\label{}
    P_1(\Phi) = \int_0^\infty d\Theta A(\Phi,\Theta)P_0(\Theta)
\end{equation}
where $A(\Phi,\Theta)$ is the Green function associated to the
evolution of probabilities in $\Phi$ space. That is
$A(\Phi,\Theta)$ represents the evolved at time $t_f$ of a Dirac
delta centered around $\Theta$ at time $t_i$. If we denote the
time evolution operator that evolves probabilities in $\Phi$ space
from time $t_i$ to time $t_f$ as $\mathscr{U}$, $A$ is defined as:
\begin{equation}\label{eq:def-A}
    A(\Phi,\Theta) = \mathscr{U} \delta(\Phi-\Theta)
\end{equation}
The function $A(\Phi,\Theta)$ is the classical counterpart of the
transition probability $|a_{kn}|^2$. Evidently, thanks to the
classical adiabatic theorem we have for an adiabatic switching:
\begin{equation}\label{}
    A(\Phi,\Theta) = \delta(\Phi-\Theta)
\end{equation}
This is the classical counterpart of Eq. (\ref{eq:a-kn=delta-kn}).
For non adiabatic switching we expect $A(\Phi,\Theta)$, considered
as a function of $\Phi$, to be bell-shaped with some finite width.
The problem of determining the shape of $A$ has been studied by
\cite{Jarz92}, who proved that, within the second order of
adiabatic perturbation theory, $A$ actually drifts and diffuses
according to an effective Fokker-Planck equation. Since
$A(\Phi,\Theta)$ represents a probability distribution function in
$\Phi$ space, it satisfies:
\begin{equation}\label{}
    A(\Phi,\Theta) \geq 0
\end{equation}
and
\begin{equation}\label{}
    \int_0^\infty d\Phi A(\Phi,\Theta) = 1
\end{equation}
Using Liouville's Theorem it is also possible to prove that:
\begin{equation}\label{}
    \int_0^\infty d\Theta A(\Phi,\Theta)= 1.
\end{equation}
The analogy with the quantum case has been completely established
now, and the proof of the Entropy Principle follows by repeating
the same steps. The requirement on the initial distribution
$P_0(\Phi)$ is that it be a decreasing function of $\Phi$. Since
$\Phi_0(E)$ is increasing, this requirement translates into the
requirement that $p_0(E)$ be a decreasing function of $E$.

\section{\label{sec:initial-condition}The role of the initial equilibrium}
\begin{figure}
\includegraphics[width=13cm]{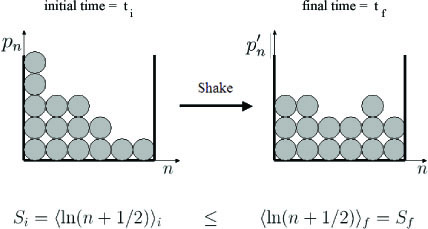}
  \caption{Visual representation of the effect of a non adiabatic perturbation on a quantum system which is initially described by decreasing
  probabilities. After shaking, the initial accumulation towards the left would flatten out,
  and the average value of $\ln(n+1/2)$ (i.e. the entropy) would increase.}
\label{fig:shake}
\end{figure}

Inequality (\ref{eq:Entropy-Increase}) holds as a direct
consequence of the time reversal symmetric microscopic laws of
quantum or classical mechanics. As such, it does not entail any
arrow of time. The reason for the emergence of the $\geq$ sign in
Eq. (\ref{eq:Entropy-Increase}) should be looked for, rather, in
the fact that we have considered only a certain restricted subset
of all possible initial conditions. To explain this point it might
be useful to see our ensemble of systems as a box containing many
balls (see Fig. \ref{fig:shake}).
%------------------------------------------------------------------------------------------------------------------------------figure -----------
%-------------------------------------------------------------------------------------------------------------------------------------------------
Each ball represents an element of the ensemble. The box is
divided into labelled cells that represent the quantum states. The
cell most close to the left wall is state with $n=0$, its right
neighbor cell is the state $n=1$ and so forth. At time $t_i$,  the
balls are distributed in the box according to some probability
$p_n$. We can see the time dependent perturbation acted on the
system as the action of shaking the box. The effect of the shaking
is that of flattening out the initial distribution. Thus if
initially we had some accumulation of balls towards the left side
of the box, we expect the final state to be more flat. If we look
at the average value of $n$ or any other increasing function of
$n$, like for example $\ln(n+1/2)$, we would record an increase of
such values. This is a mere consequence of the fact that initially
we had an accumulation towards the left. If initially we have had
an accumulation towards the right, again the shaking would flatten
out the distribution, but this time we would see a decrease of the
average value of $n$ and of $\ln (n+1/2)$. If instead the initial
distribution were flat, we would see no change in those
quantities. Indeed it is quite easy to see that the sign of
inequality (\ref{eq:Entropy-Increase}) would be reversed if an
increasing ordering of initial probability is assumed. Therefore,
for such subset of the set of all possible initial distributions,
we actually have a Law of Entropy Decrease! This reflects the fact
that there is no asymmetry in the time evolution of the Volume
Entropy operator. Thus, in principle, it should be possible to
observe a decrease of entropy if the initial equilibrium would be
given by an \emph{inverted population}. In other words, we should
be able to observe an inverted Second Law of Thermodynamics in
\emph{Negative Temperature} systems. Indeed experimental evidence
of this exists since the very pioneering works of Pound, Purcell
and Ramsey on spin systems
\citep{Pound51,Purcell51,Ramsey51,Ramsey56}. They observed that
\begin{quote}
``when a negative temperature spin system was subjected to
resonance radiation, more radiant energy was given off by the spin
system than was absorbed \citep{Ramsey56}''
\end{quote}
This means that it is possible to extract work from a negative
temperature system by means of a cyclic transformation. In other
words, for negative temperature systems we already have
experimental evidence that:
\begin{equation}\label{}
    W \leq 0
\end{equation}
Because of the \emph{equivalence} of the Minimal Work Principle
and the Entropy Principle, in this case we would also have:
\begin{equation}\label{}
    S_f \leq S_i
\end{equation}
The fact that the Law of Entropy Increase is overwhelmingly more
often observed than its symmetrical Law of Entropy Decrease is a
consequence of the fact that positive temperatures are
overwhelmingly more common than negative ones. The former in fact
is the \emph{natural} state of matter, whereas the second can only
be created \emph{artificially} and only in few very special cases.
\cite{Ramsey56} already pointed out that very strict conditions
must be met for a system to be capable of negative temperatures:
(a) the system must be at equilibrium (b) there must be an upper
limit in the Hamiltonian's spectrum (c) the system must be
thermally isolated from the environment. The second requirement is
very restrictive as most systems have an unbounded kinetic energy
term in the Hamiltonian\footnote{See \citep{Mosk05} for a recent
and interesting example of negative kinetic temperature, though.}.
Also the requirement (c) is restrictive in the sense that thermal
insulation can be achieved only approximately and for a certain
amount of time. On the contrary the inevitable thermal contact of
our system with its environment would eventually bring it to the
monotonic decreasing Gibbs state
\begin{equation}\label{}
 p_i =Z^{-1}e^{-\beta \varepsilon_i},
\end{equation}
The latter describes the natural state of matter, and as such is
the \emph{inevitable} initial condition of any thermodynamic
experiment\footnote{If we consider that in experiments on negative
temperature systems one has first to create an inverted population
from a natural one, we will see that indeed the total entropy
change would be positive. The entropy spent to create the inverted
population is larger than that gained back when applying the
resonant radiation.}. Thus the time asymmetry of the laws of
thermodynamics arises at the level of the initial thermal
equilibrium, rather than in the Second Law itself. This seems to
be in agreement with the view expressed by \citet{Brown01},
according to which the Second Law does not entail any time
asymmetry. The origin of the arrow of time should be looked for,
instead, in the Minus First Law of thermodynamics, namely the
\emph{Equilibrium Principle} \citep{Brown01}.

\section{\label{sec:conclusion}Conclusion}
Adopting an approach similar to those adopted previously to prove
the Thomson's formulation of the Second Law, here we have proved
the Entropy Principle on the basis of Quantum Mechanics, for
initial conditions characterized by decreasing probabilities. This
completes a programme devoted to the study of the mechanical
foundations of Thermodynamics initiated with a previous work
\citep{Campisi05}. That work addressed the equilibrium part of the
Second Law, whereas the present one addresses the out of
equilibrium part. The key tool of investigation adopted in both
studies is the Volume Entropy. Here we have compared it to other
quantum entropies and shown that it proves more effective in
addressing the Entropy Principle. We discussed the equivalence of
the Entropy Principle and Thomson's Principle and have seen that
the Entropy Principle can be proved classically as well. The
apparent time asymmetry expressed by the Second Law stems from the
fact that initial decreasing probabilities are overwhelmingly more
common and natural in ordinary experimental set-ups than
increasing ones. Indeed a Reversed Entropy Principle can be
observed in artificially prepared negative temperature systems.

\bibliographystyle{elsart-harv}

%\bibliography{thebibliography}% Produces the bibliography via BibTeX.

\end{document}